\documentclass{article}
\usepackage{amssymb}
\usepackage{eucal}
\usepackage{amsmath}
\usepackage{amsthm}
\usepackage{graphicx}
\usepackage{float}
\usepackage{framed}
\usepackage{hyperref}
\usepackage{stackrel}
\usepackage{amscd,amssymb,amsthm,graphicx,multicol,eucal,amsmath,framed,hyperref,color,enumerate,epstopdf}

\usepackage{bm}
\usepackage{cite}
\usepackage{authblk}

\newcommand{\R}{{\mathbb  R}}  \numberwithin{equation}{section} \newtheorem{thm}{\bf
Theorem}[section]
 \newtheorem{prop}[thm]{\bf Proposition} \newtheorem{cor}[thm]{\bf Corollary}
  \theoremstyle{remark}
\newtheorem{rem}{\bf
Remark}[section]  
\DeclareMathOperator{\tr}{tr}

\begin{document}

\title{\Large \bf Characterization of the critical points for the free energy of a Cosserat problem}

 \author{Petre Birtea, Ioan Ca\c su and Dan Com\u{a}nescu \\ {\small Department of Mathematics, West University of Timi\c soara}\\ {\small Bd. V.
P\^ arvan,
No 4, 300223 Timi\c soara, Rom\^ania}\\ {\small petre.birtea@e-uvt.ro, ioan.casu@e-uvt.ro, dan.comanescu@e-uvt.ro}\\ }
\date{ }

\maketitle

\begin{abstract}
Using the embedded gradient vector field method we explicitly compute the list of critical points of the free energy for a Cosserat body model. We also formulate necessary and sufficient conditions for critical points in the abstract case of the special orthogonal group $SO(n)$. Each critical point is then characterized using an explicit formula for the Hessian operator of a cost function defined on the orthogonal group. We also give a positive answer to an open question posed in L. Borisov, A. Fischle, P. Neff, \textit{Optimality of the relaxed polar factors by a characterization of the set of real square roots of real symmetric matrices}, ZAMM (2019), namely if all local minima of the optimization problem are global minima. We point out a few examples with physical relevance,
in contrast to some theoretical (mathematical) situations that do not hold such a relevance.
\end{abstract}

{\bf MSC}: 74A30, 74A60, 74G05, 74G65, 74N15, 53B21.

{\bf Keywords}: Optimization, Hessian operator, Constraint manifold, Orthogonal group, Cosserat body.

\maketitle

\section{Introduction}

In 1909 the paper \cite{cosserat-1909} of E. Cosserat and F. Cosserat presents an extension of the elastic continuum theory. In present days the theory of Cosserat brothers is considered a particular case of the theory of complex continuum. In our paper we search for "small" rotations that minimize the free energy of a linear model for a Cosserat body, energy that is obtained in the hypotheses of isochoric homogeneous deformation and homogeneous microrotations. Details regarding the physical model are presented in Section 2, following the papers \cite{eringen, neff-fischle-munch, teodorescu, neff-lankeit-madeo, bohmer-neff-seymenoglu, fischle-neff-raabe}. There is a rich literature on nonlinear models of a Cosserat body, see \cite{neff-munch, bohmer-lee-neff}.

In Section 3 we extend the initial optimization problem for the linear Cosserat model,  that is defined on $SO(3)$, to the general case of the special orthogonal group $SO(n)$. We start by recalling the embedded gradient vector field method, \cite{Birtea-Comanescu-Hessian, 5-electron, birtea-casu-comanescu, birtea-casu-comanescu-second}, and apply it to obtain necessary and sufficient conditions for a rotation to be a critical point of the generalized free energy. 

In the last section, using the embedded gradient vector field method, we compute explicitly all the critical points of the free energy for the considered model in the cases $n=2$ and $n=3$. We recover the critical points previously found in \cite{borisov-fischle-neff}, where an ingenious algebraic technique is used. 
In \cite{neff-siaga} the same problem has been addressed and solved for the more general case of the group $O(n)$.
Using the explicit formula for the Hessian operator of a cost function defined on a constraint manifold \eqref{Hessian-general}, we completely characterize the above mentioned list of critical points. In paper \cite{borisov-fischle-neff} the question if all local minima are global minima has been left open; in our paper we give a positive answer to this conjecture. An interesting aspect is that, nevertheless, we find local maxima that are not global maxima. We underline that our analysis of the critical points is performed in all possible cases, including those when the singular values of the reduced deformation gradient \eqref{reduced-deformation-gradient} are not distinct. We also emphasize that there exist mathematical situations which have no physical relevance.

\section{Isochoric linear model for a Cosserat elastic body}

A Cosserat elastic body (or a micropolar elastic body)  consists of infinitesimal rigid bodies and the state of deformation can be characterized by {\it the deformation vector} $\boldsymbol{\varphi}$ and {\it the Cosserat microrotation} ${A}\in \mathfrak{so}(3)$ (skew-symmetric $3\times 3$ matrix). In some situations the microrotation is described by ${R}=\exp({A})\in SO(3)$.

By using the above mathematical objects one can introduce {\it the displacement vector} ${\bf u}({\bf X})=\varphi({\bf X})-{\bf X}$, {\it the deformation gradient} $F:=\nabla \boldsymbol{\varphi}$, {\it the infinitesimal non-symmetric first Cosserat strain tensor}  ${\boldsymbol{\epsilon}}=\nabla {\bf u}-{A}=F-{\mathbb{I}}_3-{A}$, and {\it the micropolar curvature tensor} $\nabla \text{axl}({A})$. 

The state of stress of a Cosserat body is described by {\it the stress tensor} $\boldsymbol{\sigma}$ and {\it the couple-stress tensor} ${\bf m}$.
The free energy is 
\begin{equation}\label{free-energy}
W=\frac{1}{2}\left<\boldsymbol{\sigma},{\boldsymbol{\epsilon}}\right>+\frac{1}{2}\left<{\bf m},\nabla \text{axl}({A})\right>=\frac{1}{2}\tr(\boldsymbol{\sigma}^T{\boldsymbol{\epsilon}})+\frac{1}{2}\tr({\bf m}^T\nabla \text{axl}({A})).
\end{equation}

For the linear isotropic elastic model the constitutive laws, see \cite{eringen} and \cite{teodorescu} are given by\footnote{If $X\in \mathcal{M}_n(\R)$, then $\text{sym}(X)=\frac{1}{2}(X+X^T)$ and $\text{skew}(X)=\frac{1}{2}(X-X^T)$.\\\\
 The map $\text{axl}:\mathfrak{so}(3)\rightarrow \R^3$ is the canonical identification of $\mathfrak{so}(3)$ with $\R^3$. It is defined the identity $\text{axl}(X)\times {\bf x}=X{\bf x}$ for all ${\bf x}\in \R^3$ and $''\times''$ is the cross-product on $\R^3$. By using components we have 
$\text{axl}\begin{pmatrix}
0 & -x_3 & x_2 \\
x_3 & 0 & -x_1 \\
-x_2 & x_1 & 0
\end{pmatrix}=
\begin{pmatrix}
x_1 \\ x_2 \\ x_3
\end{pmatrix}$.}
\begin{align}
\boldsymbol{\sigma}= & \lambda\tr({\boldsymbol{\epsilon}}) \mathbb{I}_3+2\mu \text{sym}({\boldsymbol{\epsilon}})+2\mu_c \text{skew}({\boldsymbol{\epsilon}}) \\
{\bf m}= & \alpha \tr(\nabla \text{axl}({A})) \mathbb{I}_3+(\beta+\gamma) \text{sym}(\nabla \text{axl}({A}))+(\beta-\gamma) \text{skew}(\nabla \text{axl}({A})), 
\end{align}
where $\lambda, \mu$  are the {\it elastic Lam\'e moduli}, $\mu_c$  is {\it the Cosserat modulus}, $\alpha, \beta, \gamma$ are the micropolar curvature moduli.

For a linear elastic Cosserat body the free energy \eqref{free-energy} takes the form (see  \cite{eringen}, \cite{teodorescu})

\begin{equation}
W=W_{mp}+W_{curv},
\end{equation}
where
\begin{equation}
W_{mp}=\mu\|\text{sym}({\boldsymbol{\epsilon}})\|^2+\mu_c\|\text{skew}({\boldsymbol{\epsilon}})\|^2+\frac{\lambda}{2}(\tr({\boldsymbol{\epsilon}}))^2
\end{equation}
is {\it the strain energy} and 

\begin{equation}
W_{curv}=\frac{\beta+\gamma}{2}\|\text{sym}(\nabla \text{axl}({A}))\|^2+\frac{\beta-\gamma}{2}\|\text{skew}(\nabla \text{axl}({A}))\|^2+\frac{\alpha}{2}(\tr(\nabla \text{axl}({A})))^2
\end{equation}
is {\it the curvature energy}. 

For a {\it homogeneous deformation and microrotation} the deformation gradient $F$ and  the Cosserat microrotation ${A}$ do not depend on the particles and consequently, the micropolar curvature tensor is $\nabla \text{axl}({A})=0$. In this case the couple-stress tensor is ${\bf m}=0$ and the curvature energy is $W_{curv}=0$.

\subsection*{Small displacements and small rotations}
We can write
$${R}^T=(\exp({A}))^T=\mathbb{I}_3-{A}+{A}^2-\dots$$
and consequently,
\begin{equation}
{R}^TF-\mathbb{I}_3=(\mathbb{I}_3-{A}+{A}^2-\dots)(\nabla {\bf u}+\mathbb{I}_3)-\mathbb{I}_3=\nabla {\bf u}-{A}-(\nabla {\bf u}){A}+{A}^2-\dots=\boldsymbol{\epsilon}-{A}\nabla {\bf u}+{A}^2-\dots
\end{equation}

We consider, as in \cite{bohmer-neff-seymenoglu}, small displacements $\|\nabla u\| \ll 1$ and small rotations $\|A\|\ll 1$; more precisely we neglect the terms ${A}^k(\nabla {\bf u})$ with $k\geq 1$ and ${A}^q$ with $q\geq 2$.
In this situation we obtain
\begin{equation}
\boldsymbol{\epsilon}={R}^TF-\mathbb{I}_3.
\end{equation}

An isochoric deformation is characterized by the equality 
\begin{equation}
\det F= 1.
\end{equation}
For small displacements and small rotations we have
$$\det F=\det (\bm{\epsilon}+\mathbb{I}_3+{A})=1+{\epsilon}_{11}+{\epsilon}_{22}+{\epsilon}_{33}+\text{h.o.t}\approx 1+{\epsilon}_{11}+{\epsilon}_{22}+{\epsilon}_{33}=1+\tr (\bm{\epsilon}).$$
and consequently the condition of an isochoric deformation is given by the equality
\begin{equation}
\tr (\bm{\epsilon})=0.
\end{equation}

Synthesizing the above discussion, the free energy in the linear isotropic model for a Cosserat body, in the hypotheses of isochoric homogeneous deformation and homogeneous microrotation, becomes
\begin{equation}\label{free-energy-SO(3)}
W:=\widetilde{W}_{\mu,\mu_c}=\mu\|\hbox{sym}(R^TF-\mathbb{I}_3)\|^2+\mu_c\|\hbox{skew}(R^TF-\mathbb{I}_3)\|^2,
\end{equation}
with $\widetilde{W}_{\mu,\mu_c}(~\cdot ~;F):SO(3)\to \mathbb{R}$, and $\mu>0$ and $\mu_c\geq 0$.

Following the principle of least action or ''action euclidienne'', see \cite{cosserat-1909}, pp. 156 and \cite{neff-2006}), in the quasi-static case and the above hypotheses, we search for ''small'' rotations that minimize the free energy $\widetilde{W}_{\mu,\mu_c}$, i.e.
\begin{equation}
\stackrel[R\in SO(3)]{}{\text{\normalsize argmin}}\widetilde{W}_{\mu,\mu_c}(R;F).
\end{equation} 

The above optimization problem will be further reduced with respect to the coefficients $\mu$, $\mu_c$, see \cite{fischle-neff-2017}, to case $\mu=1$, $\mu_c=0$.

\section{Optimization of free energy on $SO(n)$}

We remind {\bf the embedded gradient vector field method} for optimization of cost functions defined on constraint manifolds, see \cite{Birtea-Comanescu-Hessian, 5-electron, birtea-casu-comanescu, birtea-casu-comanescu-second}. 
If $\mathcal{S}\subset \mathfrak{M}$ is a submanifold of a Riemannian manifold $(\mathfrak{M},{\bf g})$ described by a set of constraint functions, i.e. $\mathcal{S}={\bf F}^{-1}(c_0)$, where ${\bf F}=(F_1,\dots,F_k):\mathfrak{M}\rightarrow \R^k$ is a smooth map and $c_0\in \R^k$ is a regular value of ${\bf F}$, then the manifold $\mathcal{S}$ becomes a Riemannian manifold endowed with the induced metric ${\bf g}_{_{ind}}$. 
In optimization problems we need, in general, to compute the gradient vector field and the Hessian operator of a smooth cost function $\widetilde{G}:(\mathcal{S},{\bf g}_{_{ind}})\rightarrow \R$. 

Let $G:(\mathfrak{M},{\bf g})\rightarrow \R$ be a smooth prolongation of $\widetilde{G}$. In \cite{birtea-comanescu, Birtea-Comanescu-Hessian, 5-electron}, it has been proved that
\begin{equation}\label{partial-G-134}
\nabla _{{\bf g}_{_{ind}}}{G}_{|{\bf F}^{-1}(c)}(s)=\partial_{\bf g} G(s),\,\,\,\forall s\in {\bf F}^{-1}(c),\,\,\,\text{and}\,\,c\,\,\text{an arbitrary regular value},
\end{equation}
where $\partial_{\bf g} G$ is defined on the open set of regular points $\mathfrak{M}^{reg}\subset \mathfrak{M}$ of the constraint function, and it is the unique vector field that is tangent to the foliation generated by ${\bf F}$ having property \eqref{partial-G-134}. {The embedded gradient vector field}  is given by the following formula:
\begin{equation}\label{partial-G-general}
\partial_{\bf g} G(s)=\nabla_{\bf g} G(s)-\sum\limits_{i=1}^k\sigma_{\bf g}^{i}(s)\nabla_{\bf g}F_i(s),\,\,\,\forall\,s\in \mathfrak{M}^{reg}.
\end{equation}
The Lagrange multiplier functions $\sigma_{\bf g}^{i}:\mathfrak{M}^{reg}\rightarrow \R$ are defined by the formula

\begin{equation*}\label{sigma-101}
\sigma^i_{\bf g}(s):=\frac{\det \left(\text{Gram}_{(F_1,\ldots ,F_{i-1},G, F_{i+1},\dots,F_k)}^{(F_1,\ldots , F_{i-1},F_i, F_{i+1},...,F_k)}(s)\right)}{\det\left(\text{Gram}_{(F_1,\ldots ,F_k)}^{(F_1,\ldots ,F_k)}(s)\right)},
\end{equation*}
where
\begin{equation*}\label{sigma}
\text{Gram}_{(g_1,...,g_s)}^{(f_1,...,f_r)}:=\left[%
\begin{array}{cccc}
  {\bf g}(\nabla_{\bf g} g_1,\nabla_{\bf g}f_{1}) & ... & {\bf g}(\nabla_{\bf g} g_s,\nabla_{\bf g} f_{1}) \\
  \vdots & \ddots & \vdots \\
  {\bf g}(\nabla_{\bf g} g_1,\nabla_{\bf g}f_r) & ... & {\bf g}(\nabla_{\bf g} g_s,\nabla_{\bf g} f_r) 
\end{array}%
\right].
\end{equation*}
As mentioned in \cite{Birtea-Comanescu-Hessian} in a critical point $s_0\in \mathcal{S}$ of $\widetilde{G}$  the numbers $\sigma^i_{\bf g}(s_0)$ coincide with the classical Lagrange multipliers.

Also, in \cite{Birtea-Comanescu-Hessian, 5-electron} it has been proved that 
\begin{equation}\label{Hessian-general}
\text{Hess}_{{\bf g}_{_{ind}}}\, \widetilde{G}(s) =\left(\text{Hess}_{\bf g}\,G(s)-\sum_{i=1}^k\sigma_{\bf g}^{i}(s)\text{Hess}_{\bf g}\, F_i(s)\right)_{|T_s \mathcal{S}\times T_s \mathcal{S}},\,\,\,\forall s\in \mathcal{S}.
\end{equation}

We apply the method described above to the particular case of $SO(n)$ regarded as a submanifold  of $\mathcal{M}_{n\times n}(\R)$.
For $n\geq 2$, we consider the special orthogonal group:
$$SO(n)=\{R\in \mathcal{M}_{n\times n}(\R) \,|\,R^TR=RR^T=\mathbb{I}_n,\,\,\det{R}=1\}.$$
Denote with  ${\bf r}_1,...,{\bf r}_n\in \R^n$ the  vectors that form the columns of the matrix $R\in  \mathcal{M}_{n\times n}(\R)$.

The constraint functions defining $SO(n)$ are $F_{aa},F_{bc}:\mathcal{M}_{n\times n}({\R})\rightarrow \R$ given by:
\begin{align}
F_{aa} (R) & =\frac{1}{2}\|{\bf r}_a\|^2,\,\,1 \leq a\leq n,\nonumber \\
F_{bc} (R) & = \left<{\bf r}_b,{\bf r}_c\right>,\,\,1\leq b<c\leq n. \nonumber 
\end{align}
We have ${\bf F}:\mathcal{M}_{n\times n}({\R})\rightarrow \R^{\frac{n(n+1)}{2}}$, ${\bf F}:=\left( \dots , F_{aa},\dots ,F_{bc}, \dots\right)$, 
$$SO(n)\simeq {\bf F}^{-1}\left( \dots , \frac{1}{2},\dots ,0, \dots\right)\subset \mathcal{M}_{n\times n}({\R}).$$
More precisely, $SO(n)$ is the connected component of $ {\bf F}^{-1}\left( \dots , \frac{1}{2},\dots ,0, \dots\right)$ determined by the condition $\det(R)=1$.

In what follows we denote by ${\bf vec}(R)\subset \R^{n^2}$ the column vectorization of the matrix $U\in \mathcal{M}_{n\times n}(\R)$. Also, for a smooth function $f:\mathcal{M}_{n\times n}(\R)\rightarrow \R$ we denote 
$\nabla f(R):=\nabla (f\circ {\bf vec}^{-1})({\bf vec}(R))\in \R^{n^2}$
and 
$\text{Hess}\,f(R):=\text{Hess}\,(f\circ {\bf vec}^{-1})({\bf vec}(R))\in \mathcal{M}_{n^2\times n^2}(\R).$

The Lagrange multiplier functions for the case of $SO(n)$, see \cite{Birtea-Comanescu-Hessian}, are given by: 
\begin{equation}\label{Lagrange-multipliers-functions}
\left.\begin{array}{l}
\sigma_{aa}(R)=  \left<\nabla G(R),\nabla F_{aa}(R)\right>=\left<\displaystyle\frac{\partial G}{\partial {\bf r}_a}(R),{\bf r}_a\right>;\\ \\
\sigma_{bc}(R)= \left<\nabla G(R),\nabla F_{bc}(R)\right>=\displaystyle\frac{1}{2}\left(\left<\displaystyle\frac{\partial G}{\partial {\bf u}_c}(R),{\bf r}_b\right>+\left<\displaystyle\frac{\partial G}{\partial {\bf r}_b}(R),{\bf r}_c\right>\right),\end{array}\right.
\end{equation}
where $G:\mathcal{M}_{n\times n}({\R})\rightarrow \R$ is the extension of the cost function $\widetilde{G}:SO(n)\rightarrow \R$.

We consider the symmetric matrix
$$\Sigma(R): =\left[\sigma_{bc}(R)\right]\in \mathcal{M}_{n\times n}(\R),$$
where we define $\sigma_{cb}(R):=\sigma_{bc}(R)$ for $1\leq b<c\leq n$.
Using \eqref{Lagrange-multipliers-functions}, we have
\begin{equation}\label{Sigma}
\Sigma(R)=\frac{1}{2}\left((\nabla G(R))^T R+R^T\nabla G(R)\right).
\end{equation}

The matrix form of the embedded gradient vector field $\partial G$ is given by (see  \cite{birtea-casu-comanescu})
\begin{equation}\label{partialG}
\partial G(R)=\nabla G(R)-R\Sigma(R).
\end{equation}

A necessary and sufficient condition for $R\in SO(n)$ to be a critical point of the cost function $\widetilde{G}$ is $\partial G(R)= \mathbb{O}_n$, or equivalently: 
\begin{equation}\label{critical}
R^T\nabla G(R)=(\nabla G(R))^T R.
\end{equation}

The Hessian of the cost function $\widetilde{G}:SO(n)\rightarrow \R$ is given by, see \cite{birtea-casu-comanescu-second, Birtea-Comanescu-Hessian}:
\begin{equation}\label{Hessian}
\hbox{Hess}\,\widetilde{G}(R) =\left(\hbox{Hess}\,G(R)-{\Sigma}(R)\otimes \mathbb{I}_n\right)_{|T_R SO(n)\times T_R SO(n)}.
\end{equation}

Generalizing the free energy \eqref{free-energy-SO(3)} to the case of $SO(n)$, for $F\in GL_+(n)$, we have the cost function
$\widetilde{W}_{\mu,\mu_c}(~\cdot ~;F):SO(n)\to \mathbb{R}$ given by
$$\widetilde{W}_{\mu,\mu_c}(R;F)=\mu\|\hbox{sym}(R^TF-\mathbb{I}_n)\|^2+\mu_c\|\hbox{skew}(R^TF-\mathbb{I}_n)\|^2.$$

The optimization problem is the following:
\begin{equation}
\stackrel[R\in SO(n)]{}{\text{\normalsize argmin}}\widetilde{W}_{\mu,\mu_c}(R;F).
\end{equation}

Computing explicitly the terms of $\widetilde{W}_{\mu,\mu_c}$, we have:
\begin{align}
\|\hbox{sym}(R^TF-\mathbb{I}_n)\|^2&=\frac{1}{4}
\hbox{tr}\left((R^TF+F^TR-2\mathbb{I}_n)(R^TF+F^TR-2\mathbb{I}_n)\right)\nonumber\\
&=\frac{1}{2}\hbox{tr}(R^TFR^TF)-2\hbox{tr}(R^TF)+\frac{1}{2}\hbox{tr}(FF^T)+n\nonumber;
\end{align} 

\begin{align}
\|\hbox{skew}(R^TF-\mathbb{I}_n)\|^2&=-\frac{1}{4}
\hbox{tr}\left((R^TF-F^TR)(R^TF-F^TR)\right)\nonumber\\
&=-\frac{1}{2}\hbox{tr}(R^TFR^TF)+\frac{1}{2}\hbox{tr}(F^TF)\nonumber.
\end{align} 

It follows that
$$\widetilde{W}_{\mu,\mu_c}(R;F)=\frac{\mu-\mu_c}{2}\hbox{tr}(F^TRF^TR)-2\mu\hbox{tr}(F^TR)+\frac{\mu+\mu_c}{2}\hbox{tr}(F^TF)+\mu n.$$

We extend naturally the cost function $\widetilde{W}_{\mu,\mu_c}(\cdot ;F)$ to ${W}_{\mu,\mu_c}(\cdot ;F):\mathcal{M}_{n\times n}(\R)\rightarrow \R$.

In order to apply the condition \eqref{critical}, we need to compute the Euclidean gradient of $W_{\mu,\mu_c}$:
$$\nabla W_{\mu,\mu_c}(R;F)=(\mu-\mu_c)FR^TF-2\mu F.$$

Consequently, the necessary and sufficient condition for a rotation $R$ to be a critical point of $\widetilde{W}_{\mu,\mu_c}$ becomes
$$\nabla W_{\mu,\mu_c}(R;F) R^T=R(\nabla W_{\mu,\mu_c}(R;F))^T,$$
or equivalently
\begin{equation}\label{critical-cosserat}
(\mu-\mu_c)(FR^TFR^T-RF^TRF^T)=2\mu (FR^T-RF^T).
\end{equation}

The next result presents necessary and sufficient conditions for critical points of the cost function $\widetilde{W}_{\mu,\mu_c}$.
\begin{thm}
The following three conditions for critical points are equivalent:
\begin{itemize}
\item[(i)] A rotation $R$ is a critical point of $\widetilde{W}_{\mu,\mu_c}$ if and only if
\begin{equation*}
(\mu-\mu_c)(FR^TFR^T-RF^TRF^T)=2\mu (FR^T-RF^T).
\end{equation*}
\item[(ii)] The matrix $R=X^TF^{-T}$ is a critical point of $\widetilde{W}_{\mu,\mu_c}$ if and only if 
\begin{equation}
(\mu-\mu_c)\left(X^2-(X^T)^2\right)=2\mu(X-X^T)
\end{equation}
and 
\begin{equation}
XX^T=FF^T.
\end{equation}
\item[(iii)] The matrix $R=X^TF^{-T}$ is a critical point of $\widetilde{W}_{\mu,\mu_c}$ if and only if 
\begin{equation}\label{iii}
(X-X^T)\left((\mu-\mu_c)(X+X^T)-2\mu\mathbb{I}_n\right)=(\mu-\mu_c)[X,X^T]\footnote{$[U,V]=UV-VU$ is the Lie bracket.}.
\end{equation}
and 
\begin{equation*}
XX^T=FF^T.
\end{equation*}
\end{itemize}
\end{thm}

The above result has been previously obtained  in an equivalent formulation, using another idea, in \cite{neff-siaga, borisov-fischle-neff}.

If $X$ is a symmetric square root of $FF^T$ with $\det X>0$, then the rotation $R=X^TF^{-T}$ is a critical point of $\widetilde{W}_{\mu,\mu_c}$, since the equality \eqref{iii} is obviously satisfied. In particular, the rotation from the polar decomposition of $F$ is a critical point of $\widetilde{W}_{\mu,\mu_c}$. 

Following an ingenious observation from \cite{fischle-neff-raabe} and \cite{fischle-neff-2017}, which discusses a reduction of the parameters $\mu,\mu_c$, we have the following straightforward computation.
Making the notation
\begin{equation}\label{reduced-deformation-gradient}
\widehat{F}_{\mu,\mu_c}:=\dfrac{\mu-\mu_c}{\mu}F\in GL(n),
\end{equation}
in the case $\mu\ne \mu_c$ we obtain the relation between the free energy $\widetilde{W}_{\mu,\mu_c}(\cdot;F)$ and the reduced free energy $\widetilde{W}_{1,0}(R;\widehat{F}_{\mu,\mu_c}):=\|\hbox{sym}(R^T\widehat{F}_{\mu,\mu_c}-\mathbb{I}_n)\|^2$:
\begin{equation}\label{W-redus}
\widetilde{W}_{\mu,\mu_c}(R;F)=\frac{\mu^2}{\mu-\mu_c}\widetilde{W}_{1,0}(R;\widehat{F}_{\mu,\mu_c})+\hbox{constant}.
\end{equation}

\begin{cor}\label{critical-condition-1999} 
\begin{itemize}
\item[(i)] If $\mu=\mu_c$ (Grioli problem), then a rotation $R$ is a critical point of $\widetilde{W}_{\mu,\mu_c}(~\cdot~;F)$ if and only if
\begin{equation}\label{grioli}
FR^T=RF^T.
\end{equation}
\item[(ii)] If $\mu\ne \mu_c$,
then a rotation $R$ is a critical point of $\widetilde{W}_{\mu,\mu_c}(~\cdot~;F)$ if and only if $R$ is a critical point of $\widetilde{W}_{1,0}(~\cdot~;\widehat{F}_{\mu,\mu_c})$, or equivalently 
\begin{equation}\label{eq-puncte-critice}
\widehat{F}_{\mu,\mu_c}R^T\widehat{F}_{\mu,\mu_c}R^T-R\widehat{F}_{\mu,\mu_c}^TR\widehat{F}_{\mu,\mu_c}^T=2(\widehat{F}_{\mu,\mu_c}R^T-R\widehat{F}_{\mu,\mu_c}^T).
\end{equation}
\end{itemize}

\end{cor}

The above corollary has been previously obtained  in an equivalent formulation, using another idea, in \cite{neff-siaga, borisov-fischle-neff}.

\begin{prop} Between the critical points of $\widetilde{W}_{\mu,\mu_c}(~\cdot~;F)$ and $\widetilde{W}_{1,0}(~\cdot~;\widehat{F}_{\mu,\mu_c})$ the following relations hold.

\begin{itemize} 
\item [(i)] The two cost functions $\widetilde{W}_{\mu,\mu_c}(~\cdot~;F)$ and $\widetilde{W}_{1,0}(~\cdot~;\widehat{F}_{\mu,\mu_c})$ have the same critical points. 

\item [(ii)] If $\mu>\mu_c$, then local minima for $\widetilde{W}_{\mu,\mu_c}(~\cdot~;F)$ are local minima for $\widetilde{W}_{1,0}(~\cdot~;\widehat{F}_{\mu,\mu_c})$, local maxima for $\widetilde{W}_{\mu,\mu_c}(~\cdot~;F)$ are local maxima for $\widetilde{W}_{1,0}(~\cdot~;\widehat{F}_{\mu,\mu_c})$, and saddle points for $\widetilde{W}_{\mu,\mu_c}(~\cdot~;F)$ are saddle points for $\widetilde{W}_{1,0}(~\cdot~;\widehat{F}_{\mu,\mu_c})$.

\item [(iii)]  If $\mu<\mu_c$, then local minima for $\widetilde{W}_{\mu,\mu_c}(~\cdot~;F)$ are local maxima for $\widetilde{W}_{1,0}(~\cdot~;\widehat{F}_{\mu,\mu_c})$, local maxima for $\widetilde{W}_{\mu,\mu_c}(~\cdot~;F)$ are local minima for $\widetilde{W}_{1,0}(~\cdot~;\widehat{F}_{\mu,\mu_c})$, and saddle points for $\widetilde{W}_{\mu,\mu_c}(~\cdot~;F)$ are saddle points for $\widetilde{W}_{1,0}(~\cdot~;\widehat{F}_{\mu,\mu_c})$.
\end{itemize}
\end{prop}


We notice that if $\mu>\mu_c$, then $\widehat{F}_{\mu,\mu_c}\in GL_+(n)$ and if $\mu<\mu_c$, then $\widehat{F}_{\mu,\mu_c}\in GL_-(n)$, as opposed to $F$ that always belongs to $GL_+(n)$. \medskip

{\bf In what follows we work under the assumption 
$\boldsymbol{\mu>\mu_c}.$}\medskip

For computational reasons, it is preferable to use instead of $\widehat{F}_{\mu,\mu_c}$  the diagonal matrix formed with the singular values of $\widehat{F}_{\mu,\mu_c}$. 
Let $\widehat{F}_{\mu,\mu_c}=R_{\mu,\mu_c}S_{\mu,\mu_c}$ be the polar decomposition of $\widehat{F}_{\mu,\mu_c}$ with $R_{\mu,\mu_c}\in SO(n)$ and $S_{\mu,\mu_c}=\sqrt{\widehat{F}_{\mu,\mu_c}^T\widehat{F}_{\mu,\mu_c}}$ is a symmetric positive definite matrix. By a change of coordinates, we can render $S_{\mu,\mu_c}$ in the diagonal form $S_{\mu,\mu_c}=Q^T_{\mu,\mu_c}D_{\mu,\mu_c}Q_{\mu,\mu_c}$ with $Q_{\mu,\mu_c}\in SO(n)$ and $D_{\mu,\mu_c}$ a diagonal matrix having as entries the singular values of $\widehat{F}_{\mu,\mu_c}$ in the decreasing order.

The following equality holds:
\begin{equation}
\widetilde{W}_{1,0}(R;\widehat{F}_{\mu,\mu_c})=\widetilde{W}_{1,0}(\overline{R};D_{\mu,\mu_c}),
\end{equation}
where the relation between the rotations $R$ and $\overline{R}$ is given by
$$\overline{R}=Q_{\mu,\mu_c}R_{\mu,\mu_c}^TRQ_{\mu,\mu_c}^T.$$

\begin{rem}\label{remarca-unu}
We denote by $\lambda_1>0,\dots,\lambda_n>0$ the singular values of $\widehat{F}_{\mu,\mu_c}$; therefore, $D_{\mu,\mu_c}=\text{diag}(\lambda_1,\dots,\lambda_n)$. Using the above equalities we deduce the following formula
\begin{equation}
\prod\limits_{i=1}^n\lambda_i=\left(\frac{\mu-\mu_c}{\mu}\right)^n\det F.
\end{equation}
\end{rem}

In paper \cite{borisov-fischle-neff}, the list of the critical points for the reduced free energy $\widetilde{W}_{1,0}(\cdot ;D_{\mu,\mu_c})$ is determined proving that any rotation that is a critical point can be written in block diagonal form with blocks of dimension one and two choosing an appropriate set of coordinates. This is a powerful result that allows determining the global minima of $\widetilde{W}_{1,0}(\cdot ;D_{\mu,\mu_c})$ in any dimension. 
In \cite{neff-siaga} the same cost function has been considered for the more general case of the group $O(n)$, using a similar algebraic result that characterizes matrices with symmetric square.

\section{Characterization of critical points for $n=2$ and $n=3$}

In what follows we determine and characterize the critical points, for $n\in\{2,3\}$, using the embedded gradient vector field method presented in the previous section. 

\subsection{Case 1: $n=2$}

Let $\lambda_1\geq \lambda_2>0$ be the singular values of $\widehat{F}_{\mu,\mu_c}$, i.e. the eigenvalues of $\sqrt{\widehat{F}_{\mu,\mu_c}^T\widehat{F}_{\mu,\mu_c}}$. 
Applying the necessary and sufficient condition for critical points from Corollary \ref{critical-condition-1999}, $(ii)$, we obtain the following list of critical points for the cost function $\widetilde{W}_{1,0}(\cdot; D_{\mu,\mu_c})$: 
\begin{equation}
R^{(1)}=\mathbb{I}_2,\,\,\,R^{(2)}=-\mathbb{I}_2.
\end{equation}
If $\lambda_1+\lambda_2>2$, we have the following two additional critical points:
\begin{equation}
R^{(3)}_{\pm}=\begin{pmatrix}
\frac{2}{\lambda_1+\lambda_2} & \mp\sqrt{1-\left(\frac{2}{\lambda_1+\lambda_2}\right)^2} \\
\pm\sqrt{1-\left(\frac{2}{\lambda_1+\lambda_2}\right)^2} & \frac{2}{\lambda_1+\lambda_2}
\end{pmatrix}.
\end{equation}
The values of the cost function in the above critical points are given by:
\begin{align*}
& \widetilde{W}_{1,0}(R^{(1)};D_{\mu,\mu_c})= (\lambda_1-1)^2+(\lambda_2-1)^2 \\
& \widetilde{W}_{1,0}(R^{(2)};D_{\mu,\mu_c})= (\lambda_1+1)^2+(\lambda_2+1)^2 \\
& \widetilde{W}_{1,0}(R^{(3)}_{\pm};D_{\mu,\mu_c})= \frac{1}{2}(\lambda_1-\lambda_2)^2.
\end{align*}

The following result has been previously obtained in \cite{fischle-neff-raabe, fischle-neff-2017, borisov-fischle-neff}. 

\begin{thm} (i) The critical point $R^{(2)}$ is the global maximum for the cost function $\widetilde{W}_{1,0}(\cdot;D_{\mu,\mu_c})$.

(ii) If $\lambda_1+\lambda_2\leq 2$, then the critical point $R^{(1)}$ is the global minimum for the cost function $\widetilde{W}_{1,0}(\cdot;D_{\mu,\mu_c})$.

(iii) If $\lambda_1+\lambda_2> 2$, then the critical points $R^{(3)}_{\pm}$ are the global minima for the cost function $\widetilde{W}_{1,0}(\cdot;D_{\mu,\mu_c})$.
\end{thm}

\begin{rem}\label{remarca-doi}
However, the critical points discussed in the above theorem are ''mathematical'' critical points of the cost function $\widetilde{W}_{1,0}(\cdot;D_{\mu,\mu_c})$. To have a physical significance the condition $\det F=1$ has to be satisfied. From Remark \ref{remarca-unu} we obtain the supplementary inequality constraint $\lambda_1\lambda_2=\left(\frac{\mu-\mu_c}{\mu}\right)^2<1$. For example, the situation $\lambda_1=\lambda_2=1$ qualifies for the above theorem $(ii)$, but violates the inequality constraint coming from the model of the physical problem.
\end{rem}
\medskip

\subsection{Case 2: $n=3$}

An optimization problem on the special orthogonal group $SO(3)$ can be lifted as an optimization problem for a cost function defined on the quaternions space, which is the sphere $S^3\subset \R^4$,  see \cite{altman}, \cite{shuster}, \cite{birtea-comanescu-popa}, and \cite{fischle-neff-ZAMM-2017-2}.  The smooth double covering map $\mathbf{P}:S^{3}\rightarrow SO(3)$,
$\mathbf{P}({\bf q})={R}^{\bf q}$ is a smooth local diffeomorphism defined by 
\begin{equation*}
{R}^{\bf q}=\left(
\begin{array}{ccc}
(q^0)^{2}+(q^{1})^{2}-(q^{2})^{2}-(q^{3})^{2} & 2(q^{1}q^{2}-q^{0}q^{3}) & 2(q^{1}q^{3}+q^{0}q^{2}) \\
2(q^{1}q^{2}+q^{0}q^{3})  & (q^0)^{2}-(q^{1})^{2}+(q^{2})^{2}-(q^{3})^{2}& 2(q^{2}q^{3}-q^{0}q^{1}) \\
2(q^{1}q^{3}-q^{0}q^{2}) & 2(q^{2}q^{3}+q^{0}q^{1}) & (q^0)^{2}-(q^{1})^{2}-(q^{2})^{2}+(q^{3})^{2}
\end{array}
\right)
\end{equation*}
and gives an identification of $S^3$ with $SO(3)$ up to antipodal points, i.e. the  unit quaternions ${\bf q}=(q^0,q^1,q^2,q^3)\in S^3\subset \mathbb{R}^{4}$ and $-{\bf q}\in S^3\subset\mathbb{R}^{4}$ correspond to the same rotation in $SO(3)$. 

The sphere $S^3$ is defined by the constraint smooth function $\mathbf{F}:\R^4\rightarrow \R$, $\mathbf{F}({\bf q})=\frac{1}{2}\|{\bf q}\|^2$. 
We lift the reduced free energy $\widetilde{W}_{1,0}(\cdot;D_{\mu,\mu_c})$ to the cost function $\widetilde{G}_{1,0}:S^3\rightarrow \R$, $\widetilde{G}_{1,0}:=\widetilde{W}_{1,0}(\cdot;D_{\mu,\mu_c})\circ {\bf P}$. We denote with $G_{1,0}$ the natural extension of   $\widetilde{G}_{1,0}$ to the ambient space $\R^4$.

The formula \eqref{partial-G-general} for the embedded gradient vector field becomes
\begin{equation}
\partial G_{1,0}({\bf q})=\nabla G_{1,0}({\bf q})-\frac{\left<\nabla G_{1,0}({\bf q}),{\bf q}\right>}{\|{\bf q}\|^2}{\bf q}.
\end{equation}
The critical point of the cost function $\widetilde{G}_{1,0}$ are the points ${\bf q}\in S^3$ that are the solutions of the equation
$$\partial G_{1,0}({\bf q})={\bf 0}.$$
In order to characterize the nature of the critical points we will compute the Hessian of the cost function using the formula \eqref{Hessian-general}; more precisely 
\begin{equation}\label{Hessian-sfera}
\text{Hess}\,\widetilde{G}_{1,0}\left({\bf q}\right)=\left(\text{Hess}\,{G}_{1,0}\left({\bf q}\right)-\frac{\left<\nabla G_{1,0}({\bf q}),{\bf q}\right>}{\|{\bf q}\|^2}\mathbb{I}_4\right)_{|T_{\bf q}S^3\times T_{\bf q} S^3}.
\end{equation}

A base for the tangent vector space $T_{\bf q} S^3$ can be computed in the following way, see \cite{birtea-casu-comanescu-second}. For the point ${\bf q}$ we choose an index $j\in \{0,1,2,3\}$ such that $q^j\neq 0$. The base for $T_{\bf q} S^3$ is given by $$\{{\bf e}_i-q^i{\bf q}\,|\,i\in \{0,1,2,3\}\setminus\{j\}\},$$ where ${\bf e}_0=(1,0,0,0)$, ${\bf e}_1=(0,1,0,0),$
 ${\bf e}_2=(0,0,1,0)$, and ${\bf e}_3=(0,0,0,1)$.

Let $\lambda_1\geq \lambda_2\geq \lambda_3>0$ be the singular values of $\widehat{F}_{\mu,\mu_c}$, i.e. the eigenvalues of $\sqrt{\widehat{F}_{\mu,\mu_c}^T\widehat{F}_{\mu,\mu_c}}$.  
\medskip

\subsubsection{Subcase $\lambda_1>\lambda_2>\lambda_3$}

This case has been studied in \cite{fischle-neff-ZAMM-2017-2}, where the list of critical points has been obtained by a computer assisted method, using the Reduce package from Mathematica. The rigorous proof of this result has been given in \cite{borisov-fischle-neff}. 
We rediscover the list of critical points of $\widetilde{G}_{1,0}$ and we fully characterize them using the Hessian formula \eqref{Hessian-sfera}.

\begin{enumerate}
\item ${\bf q}^{(0)}_{\pm}=(\pm 1,0,0,0)$, 
$$\text{Spec}\left(\text{Hess}\,\widetilde{G}_{1,0}\left({\bf q}^{(0)}_{\pm}\right)\right)=\{-4(\lambda_2+\lambda_3)(\lambda_2+\lambda_3-2),-4(\lambda_1+\lambda_3)(\lambda_1+\lambda_3-2), -4(\lambda_1+\lambda_2)(\lambda_1+\lambda_2-2)\}.$$
If $\lambda_2+\lambda_3>2$, then ${\bf q}^{(0)}_{\pm}$ are local strict maxima. If $\lambda_1+\lambda_2<2$, then ${\bf q}^{(0)}_{\pm}$ are local strict minima. If $\lambda_1+\lambda_2>2>\lambda_1+\lambda_3$ or $\lambda_1+\lambda_3>2>\lambda_2+\lambda_3$, then ${\bf q}^{(0)}_{\pm}$ are saddle points. The value of the cost function is $\widetilde{G}_{1,0}\left({\bf q}^{(0)}_{\pm}\right)=(\lambda_1-1)^2+(\lambda_2-1)^2+(\lambda_3-1)^2$.

\item ${\bf q}^{(1)}_{\pm}=(0,\pm 1,0,0)$, 
$$\text{Spec}\left(\text{Hess}\,\widetilde{G}_{1,0}\left({\bf q}^{(1)}_{\pm}\right)\right)=\{-4(\lambda_2+\lambda_3)(\lambda_2+\lambda_3+2),-4(\lambda_1-\lambda_3)(\lambda_1-\lambda_3-2), -4(\lambda_1-\lambda_2)(\lambda_1-\lambda_2-2)\}.$$
If $\lambda_1-\lambda_2>2$, then ${\bf q}^{(1)}_{\pm}$ are local strict maxima. If $\lambda_1-\lambda_2<2$, then ${\bf q}^{(1)}_{\pm}$ are saddle points. The value of the cost function is $\widetilde{G}_{1,0}\left({\bf q}^{(1)}_{\pm}\right)=(\lambda_1-1)^2+(\lambda_2+1)^2+(\lambda_3+1)^2$.

\item ${\bf q}^{(2)}_{\pm}=(0,0,\pm 1,0)$, 
$$\text{Spec}\left(\text{Hess}\,\widetilde{G}_{1,0}\left({\bf q}^{(2)}_{\pm}\right)\right)=\{-4(\lambda_1+\lambda_3)(\lambda_1+\lambda_3+2),-4(\lambda_1-\lambda_2)(\lambda_1-\lambda_2+2), -4(\lambda_2-\lambda_3)(\lambda_2-\lambda_3-2)\}.$$
If $\lambda_2-\lambda_3>2$, then ${\bf q}^{(2)}_{\pm}$ are local strict maxima. If $\lambda_2-\lambda_3<2$, then ${\bf q}^{(2)}_{\pm}$ are saddle points. The value of the cost function is $\widetilde{G}_{1,0}\left({\bf q}^{(2)}_{\pm}\right)=(\lambda_1+1)^2+(\lambda_2-1)^2+(\lambda_3+1)^2$.

\item ${\bf q}^{(3)}_{\pm}=(0,0,0,\pm 1)$, 
$$\text{Spec}\left(\text{Hess}\,\widetilde{G}_{1,0}\left({\bf q}^{(3)}_{\pm}\right)\right)=\{-4(\lambda_1+\lambda_2)(\lambda_1+\lambda_2+2),-4(\lambda_1-\lambda_3)(\lambda_1-\lambda_3+2), -4(\lambda_2-\lambda_3)(\lambda_2-\lambda_3+2)\}.$$
The critical points ${\bf q}^{(3)}_{\pm}$ are local strict maxima. The value of the cost function is $\widetilde{G}_{1,0}\left({\bf q}^{(3)}_{\pm}\right)=(\lambda_1+1)^2+(\lambda_2+1)^2+(\lambda_3-1)^2$.

\item If $\lambda_2-\lambda_3>2$, then we have four critical points ${\bf q}^{(4)}_{\pm;\pm}=\left(0,0,\pm\sqrt{\frac{1}{2}+\frac{1}{\lambda_2-\lambda_3}},\pm\sqrt{\frac{1}{2}-\frac{1}{\lambda_2-\lambda_3}}\right)$.
\begin{align*}
\text{Spec} & \left(\text{Hess}\,\widetilde{G}_{1,0} \left({\bf q}^{(4)}_{\pm;\pm}\right)\right)  \\
& = \left\{-4(\lambda_1+\lambda_2)(\lambda_1+\lambda_3),-4(\lambda_1-\lambda_2)(\lambda_1-\lambda_3), \frac{2(\lambda_2-\lambda_3-2)^2(\lambda_2-\lambda_3+2)}{\lambda_2-\lambda_3}\right\}.
\end{align*}
All four critical points  ${\bf q}^{(4)}_{\pm;\pm}$ are saddle points and $\widetilde{G}_{1,0}\left( {\bf q}^{(4)}_{\pm;\pm}\right)=(\lambda_1+1)^2+\frac{1}{2}(\lambda_2+\lambda_3)^2$.

\item If $\lambda_1-\lambda_3>2$, then we have four critical points ${\bf q}^{(5)}_{\pm;\pm}=\left(0,\pm\sqrt{\frac{1}{2}+\frac{1}{\lambda_1-\lambda_3}},0,\pm\sqrt{\frac{1}{2}-\frac{1}{\lambda_1-\lambda_3}}\right)$.
\begin{align*}
\text{Spec} & \left(\text{Hess}\,\widetilde{G}_{1,0} \left({\bf q}^{(5)}_{\pm;\pm}\right)\right) \\
& = \left\{-4(\lambda_1+\lambda_2)(\lambda_2+\lambda_3),4(\lambda_1-\lambda_2)(\lambda_2-\lambda_3), \frac{2(\lambda_1-\lambda_3-2)^2(\lambda_1-\lambda_3+2)}{\lambda_1-\lambda_3}\right\}.
\end{align*}
All four critical points  ${\bf q}^{(5)}_{\pm;\pm}$ are saddle points and $\widetilde{G}_{1,0}\left( {\bf q}^{(5)}_{\pm;\pm}\right)=(\lambda_2+1)^2+\frac{1}{2}(\lambda_1+\lambda_3)^2$.

\item If $\lambda_1-\lambda_2>2$, then we have four critical points ${\bf q}^{(6)}_{\pm;\pm}=\left(0,\pm\sqrt{\frac{1}{2}+\frac{1}{\lambda_1-\lambda_2}},\pm\sqrt{\frac{1}{2}-\frac{1}{\lambda_1-\lambda_2}},0\right)$.
\begin{align*}
\text{Spec} & \left(\text{Hess}\,\widetilde{G}_{1,0} \left({\bf q}^{(6)}_{\pm;\pm}\right)\right) \\
& = \left\{-4(\lambda_1+\lambda_3)(\lambda_2+\lambda_3),-4(\lambda_1-\lambda_3)(\lambda_2-\lambda_3), \frac{2(\lambda_1-\lambda_2-2)^2(\lambda_1-\lambda_2+2)}{\lambda_1-\lambda_2}\right\}.
\end{align*}
All four critical points  ${\bf q}^{(6)}_{\pm;\pm}$ are saddle points and $\widetilde{G}_{1,0}\left( {\bf q}^{(6)}_{\pm;\pm}\right)=(\lambda_3+1)^2+\frac{1}{2}(\lambda_1+\lambda_2)^2$.

\item If $\lambda_1+\lambda_2>2$, then we have four critical points ${\bf q}^{(7)}_{\pm;\pm}=\left(\pm\sqrt{\frac{1}{2}+\frac{1}{\lambda_1+\lambda_2}},0,0,\pm\sqrt{\frac{1}{2}-\frac{1}{\lambda_1+\lambda_2}}\right)$.
\begin{align*}
\text{Spec} & \left(\text{Hess}\,\widetilde{G}_{1,0} \left({\bf q}^{(7)}_{\pm;\pm}\right)\right) \\
& = \left\{4(\lambda_2+\lambda_3)(\lambda_1-\lambda_3),4(\lambda_1+\lambda_3)(\lambda_2-\lambda_3), \frac{2(\lambda_1+\lambda_2-2)^2(\lambda_1+\lambda_2+2)}{\lambda_1+\lambda_2}\right\}.
\end{align*}
All four critical points  ${\bf q}^{(7)}_{\pm;\pm}$ are strict local minima and $\widetilde{G}_{1,0}\left( {\bf q}^{(7)}_{\pm;\pm}\right)=(\lambda_3-1)^2+\frac{1}{2}(\lambda_1-\lambda_2)^2$.

\item If $\lambda_1+\lambda_3>2$, then we have four critical points ${\bf q}^{(8)}_{\pm;\pm}=\left(\pm\sqrt{\frac{1}{2}+\frac{1}{\lambda_1+\lambda_3}},0,\pm\sqrt{\frac{1}{2}-\frac{1}{\lambda_1+\lambda_3}},0\right)$.
\begin{align*}
\text{Spec} & \left(\text{Hess}\,\widetilde{G}_{1,0} \left({\bf q}^{(8)}_{\pm;\pm}\right)\right) \\
& = \left\{4(\lambda_2+\lambda_3)(\lambda_1-\lambda_2),-4(\lambda_1+\lambda_2)(\lambda_2-\lambda_3), \frac{2(\lambda_1+\lambda_3-2)^2(\lambda_1+\lambda_3+2)}{\lambda_1+\lambda_3}\right\}.
\end{align*}
All four critical points  ${\bf q}^{(8)}_{\pm;\pm}$ are saddle points and $\widetilde{G}_{1,0}\left( {\bf q}^{(8)}_{\pm;\pm}\right)=(\lambda_2-1)^2+\frac{1}{2}(\lambda_1-\lambda_3)^2$.

\item If $\lambda_2+\lambda_3>2$, then we have four critical points ${\bf q}^{(9)}_{\pm;\pm}=\left(\pm\sqrt{\frac{1}{2}+\frac{1}{\lambda_2+\lambda_3}},\pm\sqrt{\frac{1}{2}-\frac{1}{\lambda_2+\lambda_3}},0,0\right)$.
\begin{align*}
\text{Spec} & \left(\text{Hess}\,\widetilde{G}_{1,0} \left({\bf q}^{(9)}_{\pm;\pm}\right)\right) \\
& = \left\{-4(\lambda_1+\lambda_2)(\lambda_1-\lambda_3),-4(\lambda_1+\lambda_3)(\lambda_1-\lambda_2), \frac{2(\lambda_2+\lambda_3-2)^2(\lambda_2+\lambda_3+2)}{\lambda_2+\lambda_3}\right\}.
\end{align*}
All four critical points  ${\bf q}^{(9)}_{\pm;\pm}$ are saddle points and $\widetilde{G}_{1,0}\left( {\bf q}^{(9)}_{\pm;\pm}\right)=(\lambda_1-1)^2+\frac{1}{2}(\lambda_2-\lambda_3)^2$.

\end{enumerate}

Comparing the values of the  cost function $\widetilde{G}_{1,0}$ at the critical points described above we can establish global minima and global maxima for the free energy $\widetilde{W}_{1,0}(\cdot; D_{\mu,\mu_c})$.

The following result has been previously discovered in \cite{borisov-fischle-neff,fischle-neff-ZAMM-2017-2 }.

\begin{thm}
If $\lambda_1>\lambda_2>\lambda_3$, then
\begin{enumerate} [(i)]
\item The critical points ${\bf q}^{(3)}_{\pm}$ are the only global maxima for $\widetilde{G}_{1,0}$ and consequently, the corresponding rotation 
$$R^{{\bf q}^{(3)}_{\pm}}=\left( \begin {array}{ccc} -1&0&0\\ \noalign{\medskip}0&-1&0
\\ \noalign{\medskip}0&0&1\end {array} \right) $$
is the global maximum for the free energy $\widetilde{W}_{1,0}(\cdot; D_{\mu,\mu_c})$.

\item If $\lambda_1+\lambda_2>2$, then the critical points ${\bf q}^{(7)}_{\pm;\pm}$ are the only global minima for $\widetilde{G}_{1,0}$ and consequently, the corresponding rotations 
$$R^{{\bf q}^{(7)}_{+;+}}=R^{{\bf q}^{(7)}_{-;-}}=\begin{pmatrix}
\frac{2}{\lambda_1+\lambda_2} & -\sqrt{1-\left(\frac{2}{\lambda_1+\lambda_2} \right)^2} & 0 \\
\sqrt{1-\left(\frac{2}{\lambda_1+\lambda_2} \right)^2} & \frac{2}{\lambda_1+\lambda_2} & 0 \\
0 & 0 & 1
\end{pmatrix}$$
and
$$ R^{{\bf q}^{(7)}_{+;-}}=R^{{\bf q}^{(7)}_{-;+}}=\begin{pmatrix}
\frac{2}{\lambda_1+\lambda_2} & \sqrt{1-\left(\frac{2}{\lambda_1+\lambda_2} \right)^2} & 0 \\
-\sqrt{1-\left(\frac{2}{\lambda_1+\lambda_2} \right)^2} & \frac{2}{\lambda_1+\lambda_2} & 0 \\
0 & 0 & 1
\end{pmatrix}$$
are the global minima for the free energy $\widetilde{W}_{1,0}(\cdot; D_{\mu,\mu_c})$.

\item  If $\lambda_1+\lambda_2\leq 2$, then the critical points ${\bf q}^{(0)}_{\pm}$ are the only global minima for $\widetilde{G}_{1,0}$ and consequently, the corresponding rotation
$$ R^{{\bf q}^{(0)}_{\pm}}=\mathbb{I}_3 $$
is the global minimum for the free energy $\widetilde{W}_{1,0}(\cdot; D_{\mu,\mu_c})$.
\end{enumerate}

\end{thm}

As in Remark \ref{remarca-doi}, in order to satisfy the conditions of the model for the physical problem, the singular values $\lambda_1,\lambda_2,\lambda_3$ have to satisfy the inequality constraint $\lambda_1\lambda_2\lambda_3<1$. 

This case includes {\bf the simple shearing deformation} when 
$F=\begin{pmatrix}
1 & k & 0 \\
0 & 1 & 0 \\
0 & 0 & 1
\end{pmatrix},$ where $k$ a nonzero constant. The singular values of $\widehat{F}_{\mu,\mu_c}$ are distinct and are given by 
$$\lambda_1=\frac{\mu-\mu_c}{\mu}\sqrt{1+\frac{k^2}{2}+\frac{1}{2}\sqrt{k^2(k^2+4)}},\,\,\lambda_2=\frac{\mu-\mu_c}{\mu},\,\,\lambda_3=\frac{\mu-\mu_c}{\mu}\sqrt{1+\frac{k^2}{2}-\frac{1}{2}\sqrt{k^2(k^2+4)}}.$$

\bigskip

In what follows we determine and characterize the critical points in the cases of non-distinct singular values. Continuous families of critical points appear in these cases, first time discovered in \cite{borisov-fischle-neff}, and we characterize their nature using again the Hessian formula \eqref{Hessian-sfera}.

\subsubsection{Subcase $\lambda_1=\lambda_2>\lambda_3$}

The cost function $\widetilde{G}_{1,0}$ has the above critical points with exception of ${\bf q}^{(6)}_{\pm;\pm}$. Moreover, there exists three new continuous families of critical points as follows:
\begin{itemize}

\item [11.] ${\bf q}^{(10)}(\alpha)=(0,\cos\alpha,\sin\alpha,0)\,\,\alpha\in [0,2\pi).$
\begin{align*}
\text{Spec} & \left(\text{Hess}\,\widetilde{G}_{1,0} \left({\bf q}^{(10)}(\alpha)\right)\right)
= \left\{0,-4(\lambda_1+\lambda_3)(\lambda_1+\lambda_3+2),-4(\lambda_1-\lambda_3)(\lambda_1-\lambda_3-2)\right\}.
\end{align*}
The value of the cost function is $\widetilde{G}_{1,0}\left( {\bf q}^{(10)}(\alpha)\right)=2(\lambda_1^2+1)+(\lambda_3+1)^2$. If $\lambda_1-\lambda_3<2$, then the critical points  ${\bf q}^{(10)}(\alpha)$ are saddle points. 

\item [12.] If $\lambda_1-\lambda_3=\lambda_2-\lambda_3\geq 2$, then we have two families of continuous critical points 
$${\bf q}^{(11)}_{\pm}(\alpha)=\left(0, \sqrt{\frac{1}{2}+\frac{1}{\lambda_1-\lambda_3}}\cos\alpha, \sqrt{\frac{1}{2}+\frac{1}{\lambda_1-\lambda_3}}\sin\alpha, \pm \sqrt{\frac{1}{2}-\frac{1}{\lambda_1-\lambda_3}}\right),\,\,\alpha\in [0,2\pi).$$
\begin{align*}
\text{Spec} & \left(\text{Hess}\,\widetilde{G}_{1,0} \left({\bf q}^{(11)}_{\pm}(\alpha)\right)\right)
= \left\{0,-8\lambda_1(\lambda_1+\lambda_3),\frac{2(\lambda_1-\lambda_3+2)(\lambda_1-\lambda_3-2)^2}{\lambda_1-\lambda_3}\right\}.
\end{align*}
The critical points ${\bf q}^{(11)}_{\pm}(\alpha)$ are saddle points. The value of the cost function is $\widetilde{G}_{1,0}\left( {\bf q}^{(11)}_{\pm}(\alpha)\right)=(\lambda_1+1)^2+\frac{1}{2}(\lambda_1+\lambda_3)^2$.

\item  [13.] If $\lambda_1+\lambda_3=\lambda_2+\lambda_3\geq 2$, then we have two families of continuous critical points 
$${\bf q}^{(12)}_{\pm}(\alpha)=\left(\pm \sqrt{\frac{1}{2}+\frac{1}{\lambda_1+\lambda_3}},\sqrt{\frac{1}{2}-\frac{1}{\lambda_1+\lambda_3}}\cos\alpha, \sqrt{\frac{1}{2}-\frac{1}{\lambda_1+\lambda_3}}\sin\alpha, 0\right),\,\,\alpha\in [0,2\pi).$$
\begin{align*}
\text{Spec} & \left(\text{Hess}\,\widetilde{G}_{1,0} \left({\bf q}^{(12)}_{\pm}(\alpha)\right)\right) \\
& = \left\{0,-8\lambda_1(\lambda_1-\lambda_3),\frac{2(\lambda_1+\lambda_3+2)(\lambda_1+\lambda_3-2)(\lambda_1+\lambda_3-2+\cos^2\alpha(\lambda_1+\lambda_3+2))}{\lambda_1+\lambda_3}\right\}.
\end{align*}
The value of the cost function is $\widetilde{G}_{1,0}\left( {\bf q}^{(12)}_{\pm}(\alpha)\right)=(\lambda_1-1)^2+\frac{1}{2}(\lambda_1-\lambda_3)^2$.  If $\lambda_1+\lambda_3>2$, then the critical points  ${\bf q}^{(12)}_{\pm}(\alpha)$ are saddle points. 

\end{itemize}

\begin{thm}
If $\lambda_1=\lambda_2>\lambda_3$, then
\begin{enumerate} [(i)]
\item The critical points ${\bf q}^{(3)}_{\pm}$ are the only global maxima for $\widetilde{G}_{1,0}$ and consequently, the corresponding rotation 
$$R^{{\bf q}^{(3)}_{\pm}}=\left( \begin {array}{ccc} -1&0&0\\ \noalign{\medskip}0&-1&0
\\ \noalign{\medskip}0&0&1\end {array} \right) $$
is the global maximum for the free energy $\widetilde{W}_{1,0}(\cdot; D_{\mu,\mu_c})$.

\item If $\lambda_1>1$, then the critical points ${\bf q}^{(7)}_{\pm;\pm}$ are the only global minima for $\widetilde{G}_{1,0}$ and consequently, the corresponding rotations 
$$R^{{\bf q}^{(7)}_{+;+}}=R^{{\bf q}^{(7)}_{-;-}}=\begin{pmatrix}
\frac{1}{\lambda_1} & -\sqrt{1-\left(\frac{1}{\lambda_1} \right)^2} & 0 \\
\sqrt{1-\left(\frac{1}{\lambda_1} \right)^2} & \frac{1}{\lambda_1} & 0 \\
0 & 0 & 1
\end{pmatrix}$$
and
$$ R^{{\bf q}^{(7)}_{+;-}}=R^{{\bf q}^{(7)}_{-;+}}=\begin{pmatrix}
\frac{1}{\lambda_1} & \sqrt{1-\left(\frac{1}{\lambda_1} \right)^2} & 0 \\
-\sqrt{1-\left(\frac{1}{\lambda_1} \right)^2} & \frac{1}{\lambda_1} & 0 \\
0 & 0 & 1
\end{pmatrix}$$
are the global minima for the free energy $\widetilde{W}_{1,0}(\cdot; D_{\mu,\mu_c})$.

\item  If $\lambda_1\leq 1$, then the critical points ${\bf q}^{(0)}_{\pm}$ are the only global minima for $\widetilde{G}_{1,0}$ and consequently, the corresponding rotation
$$ R^{{\bf q}^{(0)}_{\pm}}=\mathbb{I}_3 $$
is the global minimum for the free energy $\widetilde{W}_{1,0}(\cdot; D_{\mu,\mu_c})$.
\end{enumerate}

\end{thm}

As in Remark \ref{remarca-doi}, in order to satisfy the conditions of the model for the physical problem, the singular values $\lambda_1=\lambda_2,\lambda_3$ have to satisfy the inequality constraint $\lambda_1^2\lambda_3<1$. 

This subcase includes the {\bf isochoric equi-biaxial stretch} when $F=\begin{pmatrix}
k & 0 & 0 \\
0 & k & 0 \\
0 & 0 & \frac{1}{k^2}
\end{pmatrix},$ where $k>1$. The singular values of $\widehat{F}_{\mu,\mu_c}$ are 
$$\lambda_1=\lambda_2=\frac{\mu-\mu_c}{\mu}\cdot k,\,\,\lambda_3=\frac{\mu-\mu_c}{\mu}\cdot \frac{1}{k^2}.$$

\subsubsection{Subcase $\lambda_1>\lambda_2=\lambda_3$}

The cost function $\widetilde{G}_{1,0}$ has the above critical points with exception of ${\bf q}^{(4)}_{\pm;\pm}$. Moreover, there exists three new continuous families of critical points as follows:
\begin{itemize}

\item [14.] ${\bf q}^{(13)}(\alpha)=(0,0,\cos\alpha,\sin\alpha)\,\,\alpha\in [0,2\pi).$
\begin{align*}
\text{Spec} & \left(\text{Hess}\,\widetilde{G}_{1,0} \left({\bf q}^{(13)}(\alpha)\right)\right)
= \left\{0,-4(\lambda_1+\lambda_2)(\lambda_1+\lambda_2+2),-4(\lambda_1-\lambda_2)(\lambda_1-\lambda_2+2)\right\}.
\end{align*}
The value of the cost function is $\widetilde{G}_{1,0}\left( {\bf q}^{(13)}(\alpha)\right)=(\lambda_1+1)^2+2(\lambda_2^2+1)$.

\item [15.] If $\lambda_1-\lambda_2=\lambda_1-\lambda_3\geq 2$, then we have two families of continuous critical points 
$${\bf q}^{(14)}_{\pm}(\alpha)=\left(0, \pm\sqrt{\frac{1}{2}+\frac{1}{\lambda_1-\lambda_2}}, \sqrt{\frac{1}{2}-\frac{1}{\lambda_1-\lambda_2}}\cos\alpha,  \sqrt{\frac{1}{2}-\frac{1}{\lambda_1-\lambda_2}}\sin\alpha\right),\,\,\alpha\in [0,2\pi).$$
\begin{align*}
\text{Spec} & \left(\text{Hess}\,\widetilde{G}_{1,0} \left({\bf q}^{(14)}_{\pm}(\alpha)\right)\right) \\
& = \left\{0,-8\lambda_2(\lambda_1+\lambda_2),\frac{2(\lambda_1-\lambda_2+2)(\lambda_1-\lambda_2-2)(\lambda_1-\lambda_2-2+\cos^2\alpha(\lambda_1-\lambda_2+2))}{\lambda_1-\lambda_2})\right\}.
\end{align*}
The value of the cost function is $\widetilde{G}_{1,0}\left( {\bf q}^{(14)}_{\pm}(\alpha)\right)=(\lambda_2+1)^2+\frac{1}{2}(\lambda_1+\lambda_2)^2$. If $\lambda_1-\lambda_2>2$, then the critical points  ${\bf q}^{(14)}_{\pm}(\alpha)$ are saddle points.

\item  [16.] If $\lambda_1+\lambda_2=\lambda_1+\lambda_3\geq 2$, then we have two families of continuous critical points 
$${\bf q}^{(15)}_{\pm}(\alpha)=\left(\pm \sqrt{\frac{1}{2}+\frac{1}{\lambda_1+\lambda_2}},0, \sqrt{\frac{1}{2}-\frac{1}{\lambda_1+\lambda_2}}\cos\alpha, \sqrt{\frac{1}{2}-\frac{1}{\lambda_1+\lambda_2}}\sin\alpha \right),\,\,\alpha\in [0,2\pi).$$
\begin{align*}
\text{Spec} & \left(\text{Hess}\,\widetilde{G}_{1,0} \left({\bf q}^{(15)}_{\pm}(\alpha)\right)\right) \\
& = \left\{0,8\lambda_2(\lambda_1-\lambda_2),\frac{2(\lambda_1+\lambda_2+2)(\lambda_1+\lambda_2-2)(\lambda_1+\lambda_2-2+\cos^2\alpha(\lambda_1+\lambda_2+2))}{\lambda_1+\lambda_2})\right\}.
\end{align*}
The value of the cost function is $\widetilde{G}_{1,0}\left( {\bf q}^{(15)}_{\pm}(\alpha)\right)=(\lambda_2-1)^2+\frac{1}{2}(\lambda_1-\lambda_2)^2$.
\end{itemize}

\begin{thm}
If $\lambda_1>\lambda_2=\lambda_3$, then
\begin{enumerate} [(i)]
\item The critical points ${\bf q}^{(2)}_{\pm}, {\bf q}^{(3)}_{\pm}, {\bf q}^{(13)}(\alpha)$ are the global maxima for $\widetilde{G}_{1,0}$ and consequently, the corresponding rotations 
$$R^{{\bf q}^{(2)}_{\pm}}=\left( \begin {array}{ccc} -1&0&0\\ \noalign{\medskip}0&1&0
\\ \noalign{\medskip}0&0&-1\end {array} \right), $$
$$R^{{\bf q}^{(3)}_{\pm}}=\left( \begin {array}{ccc} -1&0&0\\ \noalign{\medskip}0&-1&0
\\ \noalign{\medskip}0&0&1\end {array} \right), $$
and
$$R^{{\bf q}^{(13)}(\alpha)}=\left( \begin {array}{ccc} -1&0&0\\ \noalign{\medskip}0&\cos2\alpha&\sin2\alpha
\\ \noalign{\medskip}0&\sin2\alpha&-\cos 2\alpha\end {array} \right) $$
are the global maxima for the free energy $\widetilde{W}_{1,0}(\cdot; D_{\mu,\mu_c})$.

\item If $\lambda_1+\lambda_2>2$, then the critical points ${\bf q}^{(7)}_{\pm;\pm}, {\bf q}^{(8)}_{\pm;\pm}, {\bf q}^{(15)}_{\pm}(\alpha)$ are the global minima for $\widetilde{G}_{1,0}$ and consequently, the corresponding rotations 
$$R^{{\bf q}^{(7)}_{+;+}}=R^{{\bf q}^{(7)}_{-;-}}=\begin{pmatrix}
\frac{2}{\lambda_1+\lambda_2} & -\sqrt{1-\left(\frac{2}{\lambda_1+\lambda_2} \right)^2} & 0 \\
\sqrt{1-\left(\frac{2}{\lambda_1+\lambda_2} \right)^2} & \frac{2}{\lambda_1+\lambda_2} & 0 \\
0 & 0 & 1
\end{pmatrix},$$
$$ R^{{\bf q}^{(7)}_{+;-}}=R^{{\bf q}^{(7)}_{-;+}}=\begin{pmatrix}
\frac{2}{\lambda_1+\lambda_2} & \sqrt{1-\left(\frac{2}{\lambda_1+\lambda_2} \right)^2} & 0 \\
-\sqrt{1-\left(\frac{2}{\lambda_1+\lambda_2} \right)^2} & \frac{2}{\lambda_1+\lambda_2} & 0 \\
0 & 0 & 1
\end{pmatrix},$$
$$R^{{\bf q}^{(8)}_{+;+}}=R^{{\bf q}^{(8)}_{-;-}}=\begin{pmatrix}
\frac{2}{\lambda_1+\lambda_2} & 0 & \sqrt{1-\left(\frac{2}{\lambda_1+\lambda_2} \right)^2}  \\
0 & 1 & 0 \\
- \sqrt{1-\left(\frac{2}{\lambda_1+\lambda_2} \right)^2} & 0 & \frac{2}{\lambda_1+\lambda_2}
\end{pmatrix},$$
$$ R^{{\bf q}^{(8)}_{+;-}}=R^{{\bf q}^{(8)}_{-;+}}=\begin{pmatrix}
\frac{2}{\lambda_1+\lambda_2} & 0 & -\sqrt{1-\left(\frac{2}{\lambda_1+\lambda_2} \right)^2}  \\
0 & 1 & 0 \\
 \sqrt{1-\left(\frac{2}{\lambda_1+\lambda_2} \right)^2} & 0 & \frac{2}{\lambda_1+\lambda_2}
\end{pmatrix},$$
$$ R^{{\bf q}^{(15)}_{\pm}(\alpha)}=\begin{pmatrix}
\frac{2}{\lambda_1+\lambda_2} & \mp\sqrt{1-\left(\frac{2}{\lambda_1+\lambda_2} \right)^2}\cdot \sin\alpha &  \pm\sqrt{1-\left(\frac{2}{\lambda_1+\lambda_2} \right)^2}\cdot \cos\alpha\\
 \pm\sqrt{1-\left(\frac{2}{\lambda_1+\lambda_2} \right)^2}\cdot \sin\alpha & \cos^2\alpha+\frac{2}{\lambda_1+\lambda_2}\sin^2\alpha & \left(1-\frac{2}{\lambda_1+\lambda_2}\right)\sin\alpha\cos\alpha \\
\mp\sqrt{1-\left(\frac{2}{\lambda_1+\lambda_2} \right)^2}\cdot \cos\alpha &\left(1-\frac{2}{\lambda_1+\lambda_2}\right)\sin\alpha\cos\alpha & \sin^2\alpha+\frac{2}{\lambda_1+\lambda_2}\cos^2\alpha
\end{pmatrix}$$
are the global minima for the free energy $\widetilde{W}_{1,0}(\cdot; D_{\mu,\mu_c})$.

\item   If $\lambda_1+\lambda_2 \leq 2$, then the critical points ${\bf q}^{(0)}_{\pm}$ are the only global minima for $\widetilde{G}_{1,0}$ and consequently, the corresponding rotation
$$ R^{{\bf q}^{(0)}_{\pm}}=\mathbb{I}_3 $$
is the global minimum for the free energy $\widetilde{W}_{1,0}(\cdot; D_{\mu,\mu_c})$.
\end{enumerate}

\end{thm}

As in Remark \ref{remarca-doi}, in order to satisfy the conditions of the model for the physical problem, the singular values $\lambda_1,\lambda_2=\lambda_3$ have to satisfy the inequality constraint $\lambda_1\lambda_2^2<1$. 

\noindent This subcase includes the {\bf isochoric equi-biaxial stretch} when $F=\begin{pmatrix}
\frac{1}{k^2} & 0 & 0 \\
0 & k & 0 \\
0 & 0 & k
\end{pmatrix},$ where $0<k<1$. The singular values of $\widehat{F}_{\mu,\mu_c}$ are 
$$\lambda_1=\frac{\mu-\mu_c}{\mu}\cdot\frac{1}{k^2},\,\,\lambda_2=\lambda_3=\frac{\mu-\mu_c}{\mu}\cdot k.$$

\subsubsection{Subcase $\lambda_1=\lambda_2=\lambda_3=\lambda$}

In this case we find two continuous families of two dimensional critical points. The critical points of the  cost function $\widetilde{G}_{1,0}$ are the following:

\begin{itemize}

\item [17.] ${\bf q}^{(0)}_{\pm}=(\pm 1,0,0,0)$. 
\begin{align*}
\text{Spec} & \left(\text{Hess}\,\widetilde{G}_{1,0} \left({\bf q}^{(0)}_{\pm}\right)\right)
= \left\{-16\lambda(\lambda-1)\right\},
\end{align*}
where the above eigenvalue has multiplicity 3.

The value of the cost function is $\widetilde{G}_{1,0}\left( {\bf q}^{(0)}_{\pm}\right)=3(\lambda-1)^2$.

\item [18.] ${\bf q}^{(16)}=(0,q_1,q_2,q_3)$ where $q_1^2+q_2^2+q_3^2=1$. This family contains the critical points ${\bf q}^{(1)}_{\pm}$, ${\bf q}^{(2)}_{\pm}$, ${\bf q}^{(3)}_{\pm}$, ${\bf q}^{(10)}$, and ${\bf q}^{(13)}$.
\begin{align*}
\text{Spec} & \left(\text{Hess}\,\widetilde{G}_{1,0} \left({\bf q}^{(16)}\right)\right)
= \left\{0,-16\lambda(\lambda+1)\right\},
\end{align*}
where the eigenvalue 0 has multiplicity 2.

The value of the cost function is $\widetilde{G}_{1,0}\left( {\bf q}^{(16)}\right)=3\lambda^2+2\lambda+3$.

\item [19.] If $f>1$, then we have the family of critical points ${\bf q}^{(17)}_{\pm}=\left(\pm \sqrt{\frac{\lambda+1}{2\lambda}},q_1,q_2,q_3\right)$ where $q_1^2+q_2^2+q_3^2=\frac{\lambda-1}{2\lambda}$. This family contains the critical points ${\bf q}^{(7)}_{\pm;\pm}$, ${\bf q}^{(8)}_{\pm;\pm}$, ${\bf q}^{(9)}_{\pm;\pm}$, ${\bf q}^{(12)}_{\pm}$, and ${\bf q}^{(15)}_{\pm}$.
\begin{align*}
\text{Spec} & \left(\text{Hess}\,\widetilde{G}_{1,0} \left({\bf q}^{(17)}_{\pm}\right)\right)
= \left\{0,-16(\lambda+1)(q_3^2(\lambda+1)+1-\lambda)\right\},
\end{align*}
where the eigenvalue 0 has multiplicity 2.

The value of the cost function is $\widetilde{G}_{1,0}\left( {\bf q}^{(17)}_{\pm}\right)=(\lambda-1)^2$.

\end{itemize}

\begin{thm}
If $\lambda_1=\lambda_2=\lambda_3=\lambda$, then
\begin{enumerate} [(i)]
\item The critical points ${\bf q}^{(16)}$ are the global maxima for $\widetilde{G}_{1,0}$ and consequently, the corresponding rotations 
$$R^{{\bf q}^{(16)}}=\left( \begin {array}{ccc} q_{1}^2-q_2^2-q_3^2 & 2q_1q_2 & 2q_1q_3 \\ \noalign{\medskip}
2q_1q_2 & -q_{1}^2+q_2^2-q_3^2 & 2q_2q_3
\\ \noalign{\medskip}
2q_1q_3& 2q_2q_3  & -q_{1}^2-q_2^2+q_3^2\end {array} \right), $$
are the global maxima for the free energy $\widetilde{W}_{1,0}(\cdot; D_{\mu,\mu_c})$, where $q_1,q_2,q_3$ are arbitrary with the constraint $q_1^2+q_2^2+q_3^2=1$.

\item If $\lambda\leq 1$, then the critical points ${\bf q}^{(0)}_{\pm}$ are the global minima for $\widetilde{G}_{1,0}$ and consequently, the corresponding rotation 
$$ R^{{\bf q}^{(0)}_{\pm}}=\mathbb{I}_3 $$
is the global minimum for the free energy $\widetilde{W}_{1,0}(\cdot; D_{\mu,\mu_c})$.

\item   If $\lambda>1$, then the critical points ${\bf q}^{(17)}_{\pm}$ are the only global minima for $\widetilde{G}_{1,0}$ and consequently, the corresponding rotations
$$R^{{\bf q}^{(17)}_{\pm}}=\left( \begin {array}{ccc} q_{1}^2-q_2^2-q_3^2+\frac{\lambda+1}{2\lambda} & 2q_1q_2 -\sqrt{\frac{2(\lambda+1)}{\lambda}}q_3 & 2q_1q_3+\sqrt{\frac{2(\lambda+1)}{\lambda}}q_2 \\ \noalign{\medskip}
2q_1q_2+\sqrt{\frac{2(\lambda+1)}{\lambda}}q_3 & -q_{1}^2+q_2^2-q_3^2+\frac{\lambda+1}{2\lambda}  & 2q_2q_3-\sqrt{\frac{2(\lambda+1)}{\lambda}}q_1
\\ \noalign{\medskip}
2q_1q_3-\sqrt{\frac{2(\lambda+1)}{\lambda}}q_2& 2q_2q_3 +\sqrt{\frac{2(\lambda+1)}{\lambda}}q_1 & -q_{1}^2-q_2^2+q_3^2+\frac{\lambda+1}{2\lambda} \end {array} \right), $$
are the global minima for the free energy $\widetilde{W}_{1,0}(\cdot; D_{\mu,\mu_c})$, where $q_1,q_2,q_3$ are arbitrary with the constraint $q_1^2+q_2^2+q_3^2=\frac{\lambda-1}{2\lambda}$.
\end{enumerate}

\end{thm}

As in Remark \ref{remarca-doi}, in order to satisfy the conditions of the model for the physical problem, the singular value $\lambda$ has to satisfy the inequality constraint $\lambda<1$ and consequently the case $(iii)$ in the above theorem is impossible in the considered model. In all physical situations the global minimum for the free energy is the identity rotation. For $F=\mathbb{I}_3$, the case $(iii)$ from the above theorem would lead to a  body in macroscopic equilibrium having an internal rotation, but this case is excluded by the constraint 
$\lambda<1$ of the physical model.
\bigskip

\begin{rem}

In \cite{borisov-fischle-neff} it is mentioned that in the case $n=3$ it seems likely that every local minimum is automatically a global minimum. Our results confirm this conjecture in a rigorous manner. But, nevertheless, we are able to prove the existence of local maxima which are not global maxima.
\end{rem}
\bigskip

\noindent  {\bf Acknowledgment:} This work was supported by a grant of Ministery of Research and Innovation, CNCS - UEFISCDI, project number PN-III-P4-ID-PCE-2016-0165, within PNCDI III.

\end{document}